%% file: ICASSP2027_PawCT_v2.tex
\documentclass{article}
\usepackage{spconf,amsmath,graphicx,hyperref}
\usepackage{booktabs,multirow} 
\usepackage{amssymb}
\usepackage{colortbl}
\usepackage{graphicx}
\usepackage{enumitem}
\usepackage{url}
\usepackage{cite}
\usepackage[table]{xcolor}
\usepackage{tabularx}
\usepackage{array}
\definecolor{PagCTBlue}{RGB}{235,245,255}
\definecolor{conditionrow}{RGB}{235,242,250}
\definecolor{checkblue}{RGB}{31,93,153}
\definecolor{OursBlue}{RGB}{235,245,255}

\makeatletter
\let\oldthebibliography\thebibliography
\renewcommand{\thebibliography}[1]{%
  \oldthebibliography{#1}%
  \setlength{\itemsep}{0.29em}%
  \setlength{\parsep}{0pt}%
  \setlength{\parskip}{0pt}%
}
\makeatother

\title{Toward Part-Aware Choral Transcription with singing voice assignment}

\name{Hanyu Meng$^{1,4}$, Zhanhong He$^{2,4}$, Zixun Guo$^{3}$, Yaolong Ju$^{4}$\sthanks{Corresponding author, juyaolong@gbu.edu.cn.}}

\address{$^{1}$ The University of New South Wales, Sydney, Australia \\
$^{2}$ The University of Western Australia, Perth, Australia \\
$^{3}$ Center for Digital Music (C4DM), Queen Mary University of London,
London, United Kingdom \\
$^{4}$ Great Bay University, Guangdong, China}

\begin{document}
\ninept
\maketitle
\begin{abstract}
Single-instrument automatic music transcription (AMT) has advanced substantially, yet choral applications require soprano, alto, tenor, and bass (SATB) to be transcribed as separate parts. Recent note-level choral AMT instead produces a single merged note track, limiting rehearsal, education, and score reconstruction. To address this limitation, we introduce Part-aware Choral Transcription (PawCT), to our knowledge the first end-to-end neural framework that identifies active SATB parts from choral audio and transcribes each into a separate note-level track. PawCT combines part-specific onset, offset, and frame prediction with part-presence estimation, union-level supervision, and structured training targets using a range prior (RP) based on SATB pitch ranges and its ordered-continuity (OC) extension, which adds within-part melodic continuity and cross-part pitch ordering. On YouChorale, PawCT-RP-OC achieves a macro part-aware note F1 of $0.225$ at a 50-ms onset tolerance, outperforming an adapted choral baseline ($0.165$) by $36.4\%$ relative and a two-stage post-hoc assignment pipeline ($0.175$). Its part-agnostic variant, PagCT, achieves a 50-ms onset F1 of $0.382$, compared with $0.237$ for the previous state-of-the-art choral AMT model. Cross-dataset evaluations on CSD and Cantor\'ia further assess performance under dataset shift. These results demonstrate the benefit of jointly modeling note transcription and vocal-part assignment.\footnote{Code and demos are available at \url{https://hanyu-meng.github.io/Paw_Choral_AMT_Demo/}.}

\end{abstract}
\vspace{-2mm}
\section{Introduction}\label{sec:introduction}
\label{sec:intro}
\vspace{-1.5mm}
Automatic music transcription (AMT) converts audio into symbolic representations such as piano rolls, MIDI files, or notated
scores~\cite{benetos2013automatic, cumminsAudiotoScoreTranscriptionUsing2026}. Recent advances in AMT span polyphonic single-instrument music~\cite{hawthorne2018onsets,kong2021highresolution,hawthorne2021sequence,RileyGED24GAPS,LothSSGB025GOAT}, multi-instrument ensembles~\cite{gardner2022mt3,yourmt3_plus,wu2020multi,lu2023tfperceiver}, and singing voice transcription~\cite{hsu2021vocano,kum2022pseudolabel}. In particular, piano transcription has progressed rapidly through onset-conditioned and high-resolution note models~\cite{hawthorne2018onsets,kong2021highresolution} and large paired audio--MIDI datasets~\cite{hawthorne2019maestro,kong2022giantmidi}.

\begin{figure}[t]
    \centering
    \includegraphics[width=0.95\columnwidth]
    {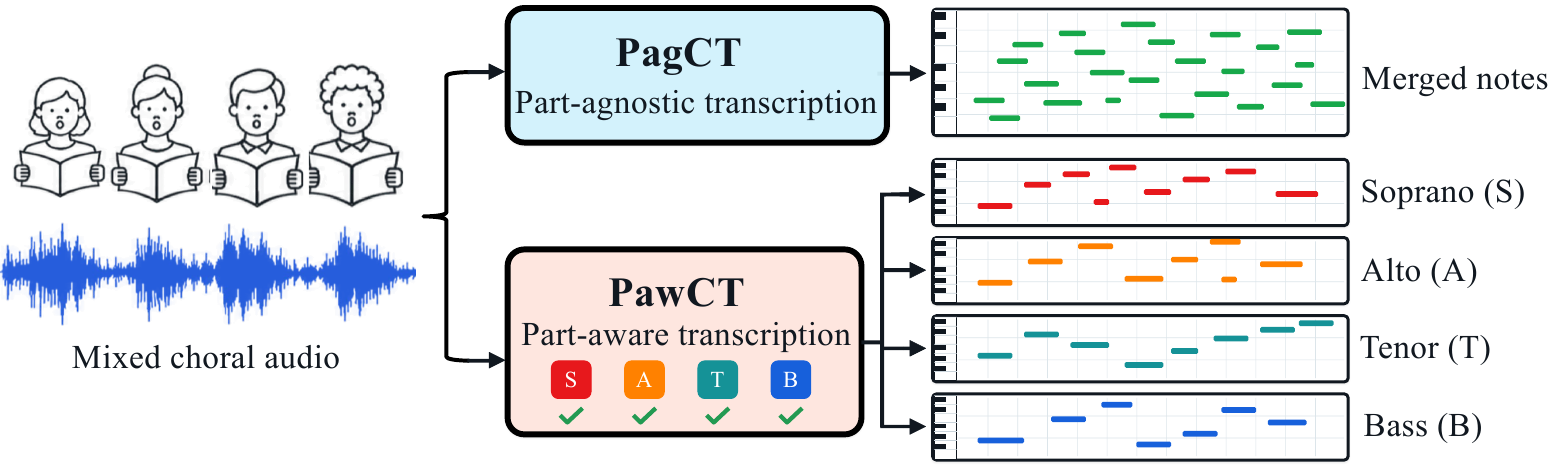}
    \vspace{-3.5mm}
    \caption{Part-agnostic versus part-aware choral transcription. PagCT produces a single merged track, whereas PawCT jointly identifies active SATB parts (green check marks) and transcribes them into separate note-level MIDI tracks.}
    \label{fig:pagct_pawct}
    \vspace{-6mm}
\end{figure}

Choral singing remains underexplored in AMT \cite{rosenzweig2020dagstuhl,yu2024note}. Choral music is commonly organized into soprano (S), alto (A), tenor (T), and bass (B) parts, which often sound simultaneously. Soft onsets and smooth pitch transitions complicate note detection, while similar vocal timbres and overlapping pitch ranges make part assignment ambiguous \cite{rosenzweig2020dagstuhl,cuesta2020multiple}. Earlier choral AMT systems estimated multi-F0 trajectories or pitch streams~\cite{cuesta2019framework,cuesta2020multiple}, with voice assignment addressed through joint acoustic--symbolic modeling~\cite{mcleod2017automatic} or neural pitch-contour models~\cite{cuesta2022voiceassignment}. These approaches focus on pitch-level representations rather than explicit note-event boundaries. Recent note-level choral AMT instead produces a single merged transcription without vocal-part labels~\cite{yu2024note}. We address the remaining gap with an end-to-end neural framework that jointly learns note transcription and active SATB-part assignment from mixed choral audio. Fig.~\ref{fig:pagct_pawct} contrasts the conventional part-agnostic setting with our part-aware formulation.

We define part-aware choral transcription as jointly identifying which SATB parts are active and transcribing each active part into a separate note-level MIDI track, yielding up to four labeled tracks. Unlike multi-instrument AMT, where instrument identity can guide track assignment \cite{gardner2022mt3,yourmt3_plus,muscriptor,lordelo2021pitch,cwitkowitz2024timbretrap}, all vocal parts belong to the same source class and provide much weaker timbral cues. To address this challenge, we introduce PawCT (\textbf{P}art-\textbf{aw}are \textbf{C}horal \textbf{T}ranscription), an end-to-end framework with the following contributions:
\begin{itemize}[
leftmargin=1.1em,
itemsep=0.2em,
topsep=0.2em,
parsep=0pt,
partopsep=0pt
]

\item \textbf{Part-aware note-level choral AMT.} We formulate joint note transcription and active SATB-part assignment from mixed audio, producing a separate note-level track for each active part.

\item \textbf{Structured end-to-end PawCT.} PawCT combines SATB-specific onset, offset, frame, and presence prediction with union-level supervision and structured range-prior (RP) and ordered-continuity (OC) training targets for part assignment.

\item \textbf{Consistent note-level gains.} On YouChorale, PagCT improves part-agnostic note F1, while PawCT-RP-OC outperforms adapted and two-stage baselines in part-aware note F1. Gains over Post-VA persist on CSD and Cantor\'ia datasets.

\end{itemize}



\vspace{-4mm}
\begin{figure*}[!t]
    \centering    
    \includegraphics[width=1.95\columnwidth]{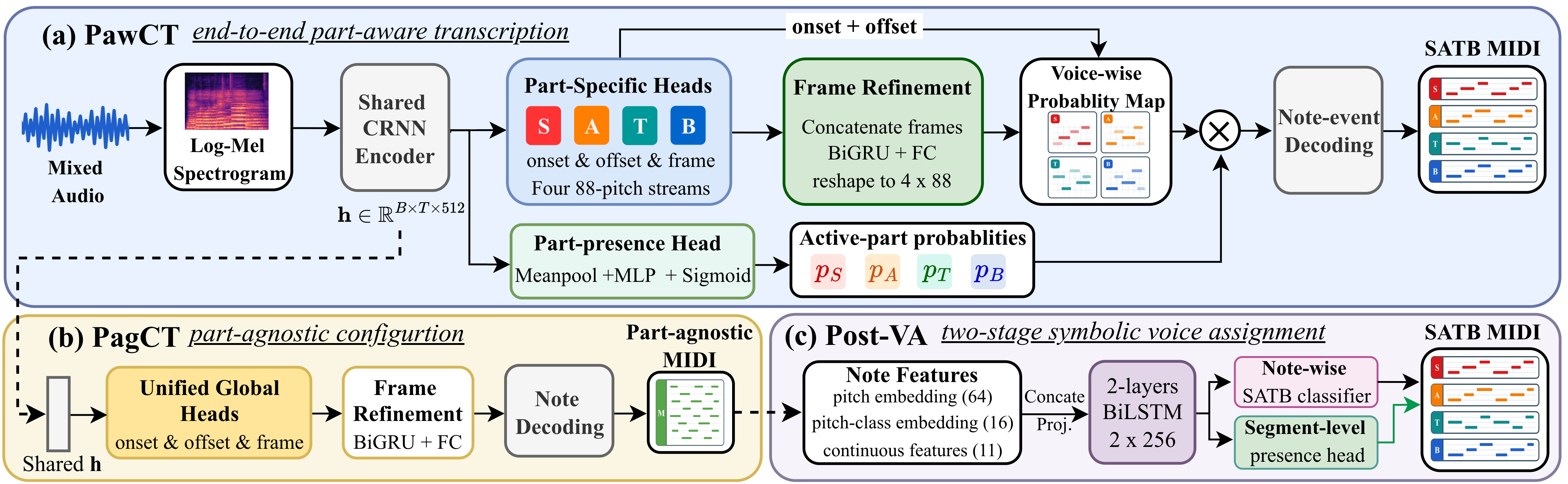}
    \vspace{-2mm}
    \caption{Two strategies for part-aware choral transcription considered in this paper. (a) The proposed PawCT jointly transcribes mixed choral audio into SATB-specific note tracks. (b)--(c) The two-stage alternative first uses PagCT to produce a merged note track and then applies Post-VA, a symbolic voice-assignment model, to assign each decoded note to a vocal part.}
    \label{fig:pct_va}
    \vspace{-5mm}
\end{figure*}

\section{Methods}
\label{sec:method}
\vspace{-2mm}
Part-aware choral AMT combines note-event detection with vocal-part assignment. Fig.~\ref{fig:pct_va} compares the proposed end-to-end PawCT model with the two-stage PagCT+Post-VA pipeline. PawCT jointly predicts SATB-specific note activity from audio using a shared convolutional recurrent neural network (CRNN) encoder and part-specific heads, while an auxiliary presence head estimates the active parts. In the two-stage pipeline, our part-agnostic variant model, PagCT, first produces a merged note track. A separately trained symbolic model, Post-VA, then assigns each decoded note to S, A, T, or B. This setup enables a comparison between joint and post-hoc voice assignment.
\vspace{-2.8mm}
\subsection{Part-aware Choral Transcription Architecture}\label{sec:pct_va_arch}
\vspace{-1mm}
PawCT (Fig.~\ref{fig:pct_va}(a)) is inspired by the high-resolution piano transcription framework~\cite{kong2021highresolution}. A log-Mel spectrogram is encoded by four convolutional blocks, each comprising two $3\times3$ convolution--batch-normalization--ReLU layers followed by average pooling along frequency. A projection layer and a two-layer bidirectional gated recurrent unit (BiGRU) with 256 units per direction then produce the frame-aligned representation $\mathbf{h}\in\mathbb{R}^{B\times T\times512}$, where $B$ and $T$ denote the batch size and number of frames.

For each vocal part $v\in\mathcal{V}=\{S,A,T,B\}$, separate sigmoid heads map $\mathbf{h}$ to onset, offset, and preliminary frame probabilities over $P=88$ MIDI pitches. The preliminary frame rolls are concatenated with the detached onset and offset rolls across all parts and refined by a one-layer BiGRU with 256 units per direction. A linear layer projects the result to $V\!\times\!P$ dimensions, where $V=4$, to obtain the refined part-wise frame probabilities. In parallel, a part-presence head mean-pools $\mathbf{h}$ over time and applies a two-layer multilayer perceptron (MLP) with dimensions 256 and 4. Its sigmoid probabilities gate the part-wise onset, offset, and frame rolls so that an inactive part produces an empty MIDI track.
\input{latex/tables/part_agnostic}
\input{latex/tables/part_aware}

\vspace{-3mm}
\subsection{Voice-Assignment Priors} \label{sec:annotation_va}
\vspace{-1mm}
Pitch proximity and temporal continuity are established cues for organizing concurrent notes into coherent musical voices \cite{mcleod2016hmm,hsiao2021learning}. In choral mixtures, however, substantial overlap among SATB pitch ranges makes either cue ambiguous on its own. We therefore introduce cumulative training-time priors: RP provides coarse register guidance, while OC builds on RP by incorporating melodic continuity, temporal overlap, and pitch-ordering constraints.

\noindent\textbf{Range prior (RP): }
RP assigns each note independently according to register. For note event $e$ with MIDI pitch $p_e$ and candidate part $v$, the assignment cost is
\begin{equation}
\vspace{-1.5mm}
\begin{aligned}
\mathcal{C}_{\mathrm{RP}}(e,v)=&\ \max(0,l_v-p_e)+\max(0,p_e-u_v) \\
&+0.25\frac{|p_e-m_v|}{12}
+\lambda_{\mathrm{part}}\mathbf{1}(v\neq v_{\mathrm{src}}),
\end{aligned}
\label{eq:rp_cost}
\end{equation}
where $[l_v,u_v]$ is the range of part $v$, $m_v=(l_v+u_v)/2$, and $v_{\mathrm{src}}$ is the canonical source label. The first two terms penalize out-of-range pitches in semitones, the third weakly favors the range center, and the final term discourages disagreement with an available source label; this term is omitted when no label is available. We use $S:[60,88]$, $A:[55,79]$, $T:[48,72]$, and $B:[40,67]$, and assign $e$ to $\hat v_e=\arg\min_{v\in\mathcal V}\mathcal{C}_{\mathrm{RP}}(e,v)$. RP imposes register structure but does not model temporal continuity.

\noindent\textbf{Ordered continuity (OC): }
OC additionally models within-part continuity and enforces pitch ordering within onset groups of up to four notes. Notes are grouped by rounded onset frame and sorted by descending pitch. For $e=(p_e,t_e^{\mathrm{on}},t_e^{\mathrm{off}})$, we define
\begin{equation}
\begin{aligned}
\mathcal{C}_{\mathrm{OC}}(e,v)=&\ \mathcal{C}_{\mathrm{RP}}(e,v)
+\lambda_{\mathrm{cont}}\frac{|p_e-\tilde p_v|}{12} \\
&+\lambda_{\mathrm{ov}}\mathbf{1}(t_e^{\mathrm{on}}<\tilde t_v^{\mathrm{off}}),
\end{aligned}
\vspace{-2mm}
\label{eq:oc_cost}
\end{equation}
where $\tilde p_v$ and $\tilde t_v^{\mathrm{off}}$ are the pitch and offset of the most recent note assigned to part $v$. The added terms penalize large within-part pitch jumps and temporally overlapping notes, respectively, and are omitted before a part has any assigned note. For onset group $G_t$ with $|G_t|\leq4$, OC selects
\begin{equation}
A_t^*=\arg\min_{A_t\in\mathcal{O}(G_t)}
\sum_{e\in G_t}\mathcal{C}_{\mathrm{OC}}\bigl(e,A_t(e)\bigr),
\label{eq:oc_assignment}
\vspace{-2mm}
\end{equation}
where $\mathcal{O}(G_t)$ contains one-to-one assignments to distinct SATB parts that preserve pitch order. Part states are updated after each group; groups containing more than four notes use independent minimum-cost assignment.

RP and OC construct the part-wise training targets from the ground-truth note annotations rather than introducing separate differentiable losses.
\vspace{-3mm}
\subsection{PawCT Training Objective} \label{sec:training_objective}
\vspace{-0.5mm}
Let $\mathcal{A}=\{\mathrm{fr},\mathrm{on},\mathrm{off}\}$ denote frame, onset, and offset activity. For each $a\in\mathcal{A}$, $\mathbf{Y}^{a}$ and $\hat{\mathbf{Y}}^{a}\in [0,1]^{B\times T\times V\times P}$ denote the resulting target and the corresponding PawCT prediction, respectively. Let $\mathcal{B}(\hat{\mathbf{Y}},\mathbf{Y},\mathbf{M})$ denote binary cross-entropy averaged over the entries selected by mask $\mathbf{M}$. The part-wise transcription loss is
\begin{equation}
\vspace{-0.5mm}
\mathcal{L}_{\mathrm{voice}}
=\sum_{a\in\mathcal A}\lambda_a^{v}
\mathcal{B}(\hat{\mathbf{Y}}^{a},\mathbf{Y}^{a},\mathbf{M}^{a}).
\label{eq:voice_loss}
\vspace{-2mm}
\end{equation}
The part-presence head is trained with $\mathcal{L}_{\mathrm{pres}}=\mathrm{BCE}(\mathbf z^p,\mathbf y^p)$, where $\mathbf z^p$ are the predicted logits and $\mathbf y^p$ indicates the active parts.

To preserve the global pitch content, we max-pool the predicted parts, $\hat{\mathbf{Y}}^{a,\cup}=\max_v\hat{\mathbf{Y}}^a_v$, and compare the result with global targets $\mathbf{Y}^{a,\cup}$ constructed from the merged ground-truth notes:
\begin{equation}
\mathcal{L}_{\mathrm{union}}
=\sum_{a\in\mathcal A}\lambda_a^{\cup}
\mathcal{B}(\hat{\mathbf{Y}}^{a,\cup},\mathbf{Y}^{a,\cup},\mathbf{M}^{a}).
\label{eq:union_loss}
\vspace{-2mm}
\end{equation}
The complete objective is
\begin{equation}
\mathcal{L}=\mathcal{L}_{\mathrm{voice}}+\mathcal{L}_{\mathrm{union}}
+\lambda_{\mathrm{pres}}\mathcal{L}_{\mathrm{pres}}.
\vspace{-1.5mm}
\label{eq:total_loss}
\end{equation}
For the ablation without union loss, all $\lambda_a^{\cup}$ are set to zero.
\vspace{-3mm}
\subsection{Part-agnostic Choral Transcription and Post-VA}
\vspace{-1mm}
\label{sec:model_variants_decoding}
PagCT (Fig.~\ref{fig:pct_va}(b)) is the part-agnostic counterpart of PawCT. Global heads estimate onset, offset, and preliminary frame rolls from the shared CRNN representation $h$, a one-layer BiGRU then refines the frame roll from their concatenation. PagCT omits the SATB-specific and part-presence heads, produces one global MIDI track, and is trained with masked BCE for the three activity types.

Post-VA tests whether SATB structure can be recovered after global transcription. Following~\cite{dong2021automatic}, it formulates VA as note-level symbolic classification. Each decoded PagCT note is represented by a 64-dimensional pitch
embedding, a 16-dimensional pitch-class embedding, and 11 continuous features projected to 64 dimensions by a two-layer MLP. Their concatenation is projected to 192 dimensions and processed by a two-layer bidirectional long short-term memory network (BiLSTM) with 256 units per direction. A four-way classifier predicts the SATB label of each note, while an auxiliary segment-level head predicts active parts. Training combines masked note-level cross-entropy, part-presence BCE, and pitch-range and within-onset crossing regularizers. The assigned notes are finally separated into four MIDI tracks.
\vspace{-3.5mm}
\subsection{Note Event Decoding}
\vspace{-1mm}
Unlike event-based methods that predict note events directly~\cite{yan2021skipping}, we decode notes from onset, offset, and frame predictions using an onset-and-frame decoder~\cite{kong2021highresolution,hawthorne2018onsets}. For each pitch, a note starts at an onset peak above the threshold and ends when offset activity exceeds the threshold, frame activity falls below the threshold, or another onset occurs. We decode each PawCT SATB output independently and PagCT once. Thresholds are selected on the YouChorale validation set and fixed during testing. The same thresholds are applied unchanged to CSD and Cantor\'ia.
\vspace{-3.5mm}
\section{Experimental Setups}\label{sec:experiment}
\vspace{-2mm}
\subsection{Data}\label{sec:experiment_data}
\vspace{-1mm}
We use YouChorale~\cite{yu2024note} as the main dataset, which contains raw audio and the notes with part labels. YouChorale contains 22 hours and 25 minutes of a cappella choral recordings, comprising 452 recordings from 261 songs with manually aligned MIDI annotations. Most recordings follow the SATB setting: 418 contain all four parts (92.48\%), while the remaining 34 contain only a subset of the SATB parts. To train the Post-VA model, we use the ground truth merged MIDI notes from the YouChorale training set together with their corresponding part labels.

We apply offline pitch transposition augmentation to the YouChorale training split. During training, each mixture and its corresponding MIDI and note annotations are jointly transposed by one to three semitones in either direction. Together with the original training data, this expands the training set by a factor of seven, yielding approximately 129 hours and 38 minutes of audio.
\vspace{-3mm}
\subsection{Model Configurations}\label{sec:model_configuration}
\vspace{-1.5mm}
We evaluate the proposed framework under both part-agnostic and part-aware settings introduced in Section~\ref{sec:method}.

For part-agnostic transcription, we compare PagCT introduced in Section~\ref{sec:model_variants_decoding} against Onsets and Frames~\cite{hawthorne2018onsets}, MT3~\cite{gardner2022mt3}, YourMT3+~\cite{yourmt3_plus}, Muscriptor~\cite{muscriptor} and the existing state-of-the-art (SOTA) choral transcription model~\cite{yu2024note}. Since~\cite{yu2024note} reports several configurations, we rerun them under the same YouChorale split, select the best configuration on the validation set, and report the corresponding test performance. We compare PagCT models trained with and without pitch-transposition augmentation, both models are evaluated on the same original, untransposed YouChorale test set.

For part-aware transcription, we compare PawCT with adapted versions of YourMT3+ as a representative multitrack AMT model, and a modified part-aware variant of~\cite{yu2024note} with SATB-specific prediction branches. We further ablate union supervision and the voice-assignment priors introduced in Section~\ref{sec:annotation_va}. The evaluated variants are PawCT without union loss, PawCT, PawCT-RP, and PawCT-RP-OC, where the final variant adds OC on top of RP. We also evaluate the two-stage PagCT~+~Post-VA pipeline as illustrated in Section~\ref{sec:model_variants_decoding}, in which Post-VA is trained on the ground-truth note sequences and part labels from the training split. All part-aware models are trained on the augmented YouChorale training set with inference thresholds of 0.03, 0.01, and 0.005 for the frame, onset, and offset predictions, respectively. 
\vspace{-3mm}
\subsection{Training Settings}
\vspace{-1mm}
Following~\cite{yu2024note}, we resample audio to 16\,kHz and compute 229-bin log-mel spectrograms with
$N_{\mathrm{FFT}}=2048$ and a 160-sample hop~\cite{kong2021highresolution}. PawCT and PagCT are trained with AdamW optimizer,
a batch size of 8, and an initial learning rate of $10^{-4}$, decayed by a factor of 0.9 every 10k iterations. Training runs for up to 200k  iterations, with early stopping after 30k iterations without validation improvement; the checkpoint with the lowest validation loss is used for evaluation. We set the RP/OC assignment-cost weights $(\lambda_{\mathrm{part}},\lambda_{\mathrm{cont}}, \lambda_{\mathrm{ov}})=(2,0.35,4)$, the part-wise frame/onset/offset loss weights to $(1,2,1)$, the corresponding union-loss weights to $(1,1,1)$, and $\lambda_{\mathrm{pres}}=0.2$. Post-VA is trained for up to 80 epochs with a batch size of 32 and early stopping patience of 10 epochs.
\input{latex/tables/cross_dataset}
\vspace{-1.8mm}
\section{Results and Discussion}
\vspace{-0.5mm}
\subsection{Evaluation Metrics}
\vspace{-0.5mm}
Following~\cite{yu2024note}, we report frame-level precision, recall, and F1 and onset-only note-level precision, recall, and F1 with 50- and 100-ms onset tolerances for part-agnostic transcription. For part-aware transcription, we report per-part and macro-average frame F1 and note F1 with a 50-ms onset tolerance. We define the voice assignment rate (VA Rate) as the percentage of part-agnostic performance retained after voice assignment: $\mathrm{VA\ Rate}=(\mathrm{F1}_{\mathrm{avg}}/\mathrm{F1}_{\mathrm{agn}})\times 100\%$, where $\mathrm{F1}_{\mathrm{avg}}$ is the macro-average across the SATB parts and $\mathrm{F1}_{\mathrm{agn}}$ is the corresponding part-agnostic F1.

\vspace{-2mm}
\subsection{Part-agnostic Transcription}
\vspace{-0.5mm}
Table~\ref{tab:youchorale_results} shows that PagCT achieves the highest note F1 at both onset tolerances. Without data augmentation, its 50-ms onset F1 is $0.333$, compared with $0.237$ for the previous choral-specific model~\cite{yu2024note}. Pitch-transposition data augmentation further raises it to $0.382$, a $61.2\%$ relative improvement over the previous model. The baselines show distinct limitations. Onsets and Frames has high precision but low recall ($0.450/0.178$), while MT3 and YourMT3+ both obtain $0.127$ note F1 at 50\,ms. MuScriptor achieves a frame F1 comparable to PagCT ($0.730/0.738$), but substantially lower note F1 at 50 and 100\,ms ($0.084/0.231$ vs. $0.382/0.573$). The high precision but low recall of Onsets and Frames suggests conservative detection of soft vocal onsets. MuScriptor's strong frame performance but lower note F1 indicates onset-timing and segmentation errors, while MT3 and YourMT3+ show limited transfer to dense vocal mixtures. Overall, accurate pitch tracking does not guarantee reliable segmentation of soft, legato notes. We therefore use augmented PagCT as the part-agnostic reference for VA Rate analysis.
\vspace{-2mm}
\subsection{Part-aware Transcription}\label{sec:partaware_results}
Table~\ref{tab:youchorale_results_part} shows that VA Rate should be interpreted together with absolute transcription accuracy. Although the adapted YourMT3+ retains $62.20\%$ of its part-agnostic note F1, its macro part-aware note F1 is only $0.079$. The adapted choral-specific model reaches $0.165$, whereas the proposed PawCT-RP-OC model achieves $0.225$, a $36.4\%$ relative improvement.

The ablations clarify the contributions of the proposed components. Union supervision raises macro note F1 from $0.190$ to $0.217$ and improves every SATB part, suggesting that mixture-level supervision helps preserve valid notes when their part assignments are uncertain. RP improves frame F1 to $0.510$ but reduces note F1 to $0.209$, indicating that vocal range helps track pitch activity but cannot resolve note assignments in overlapping vocal ranges. Adding OC raises note F1 to $0.225$ by incorporating melodic continuity and pitch ordering.

The two-stage PagCT+Post-VA pipeline obtains a macro note F1 of $0.175$ and a note-level VA Rate of $45.81\%$. PawCT-RP-OC improves these results to $0.225$ and $58.90\%$, respectively. Since Post-VA operates only on already decoded notes, it cannot recover missing or fragmented events, whereas PawCT jointly uses acoustic evidence for transcription and part assignment.

\vspace{-2mm}
\subsection{Cross-Dataset Evaluation}
As AMT systems are sensitive to recording- and corpus-level domain shifts \cite{liu2025domain}, we additionally evaluate the YouChorale-trained PagCT + Post-VA, PawCT, and PawCT-RP-OC models on CSD~\cite{csd} and Cantor\'ia~\cite{cantoria} as shown in Table~\ref{tab:cross_dataset_results}. On CSD, PawCT and PawCT-RP-OC achieve average note F1 scores of $0.161$ and $0.174$, respectively, compared with $0.118$ for the two-stage baseline. On Cantor\'ia, their scores increase to $0.271$ and $0.313$, compared with $0.115$ for PagCT + Post-VA. Compared with YouChorale, the end-to-end models perform worse on CSD but better on Cantor\'ia. The latter may partly reflect the more uniform complete-SATB structure of Cantor\'ia, although differences in recording conditions and annotation protocols may also contribute. Overall, these results support the transferability of PawCT to unseen choral datasets while motivating future work on domain adaptation and decoding calibration.
\vspace{-1mm}
\section{Conclusion}
We introduced PawCT, an end-to-end framework that identifies which SATB parts are active in choral audio and transcribes each into a separate note-level track. PawCT combines part-specific prediction, union-level supervision, and range-prior and ordered-continuity targets. On YouChorale, PagCT improves part-agnostic note F1 from $0.237$ to $0.382$, while PawCT-RP-OC achieves the highest macro part-aware note F1 among the evaluated systems, outperforming post-hoc voice assignment. Experiments on CSD and Cantor\'ia further assess generalization under dataset shift. These results demonstrate the value of jointly modeling transcription and vocal-part assignment. 

\bibliographystyle{IEEEbib}
\bibliography{refs}

\end{document}

%% file: latex/tables/part_agnostic.tex
\begin{table*}[!ht]
\caption{Part-agnostic transcription on the YouChorale test set. Following~\cite{yu2024note}, we report frame and note precision, recall, and F1, with 50- and 100-ms onset tolerances for note metrics. ``w/o Aug.'' denotes no pitch-transposition augmentation. Blue rows indicate proposed systems; bold and underlined values denote the best and second-best results, respectively.}
\vspace{1mm}
\label{tab:youchorale_results}
\centering
\small
\setlength{\tabcolsep}{2.5pt}
\renewcommand{\arraystretch}{0.88}
\setlength{\aboverulesep}{0.25ex}
\setlength{\belowrulesep}{0.25ex}

\begin{tabularx}{\textwidth}{
@{}l c *{9}{>{\centering\arraybackslash}X}@{}
}
\toprule
\multirow{2}{*}{Model}
& \multirow{2}{*}{\#Params}
& \multicolumn{3}{c}{Frame}
& \multicolumn{3}{c}{Onset (50 ms)}
& \multicolumn{3}{c}{Onset (100 ms)} \\
\cmidrule(lr){3-5}
\cmidrule(lr){6-8}
\cmidrule(lr){9-11}
& & Precision & Recall & F1
& Precision & Recall & F1
& Precision & Recall & F1 \\
\midrule

Onsets and Frames~\cite{hawthorne2018onsets}
& 6.79M
& \textbf{0.806} & 0.326 & 0.428
& \textbf{0.450} & 0.178 & 0.242
& \textbf{0.688} & 0.248 & 0.344 \\

MT3~\cite{gardner2022mt3}
& 44.7M
& 0.590 & 0.243 & 0.344
& 0.117 & 0.148 & 0.127
& 0.200 & 0.255 & 0.217 \\

YourMT3+~\cite{yourmt3_plus}
& 45.8M
& 0.529 & 0.423 & 0.457
& 0.145 & 0.119 & 0.127
& 0.252 & 0.207 & 0.221 \\

MuScriptor~\cite{muscriptor}
& 307M
& \underline{0.720} & 0.745 & 0.730
& 0.086 & 0.083 & 0.084
& 0.236 & 0.228 & 0.231 \\

Yu et al.~\cite{yu2024note}
& 4.23M
& 0.678 & 0.619 & 0.647
& 0.251 & 0.232 & 0.237
& 0.366 & 0.341 & 0.347 \\
\midrule

\rowcolor{PagCTBlue}
PagCT w/o Aug.
& 14.84M
& 0.690 & \underline{0.833} & \textbf{0.751}
& 0.346 & 0.329 & 0.333
& 0.539 & 0.506 & 0.516 \\

\rowcolor{PagCTBlue}
PagCT
& 14.84M
& 0.626 & \textbf{0.909} & \underline{0.738}
& \underline{0.398} & \textbf{0.378} & \textbf{0.382}
& \underline{0.597} & \textbf{0.569} & \textbf{0.573} \\
\bottomrule
\end{tabularx}

\vspace{-5.5mm}
\end{table*}

%% file: latex/tables/part_aware.tex
\begin{table*}[!ht]
\caption{Part-aware SATB transcription performance on the YouChorale
test set. For each voice part: Soprano, Alto, Tenor, and Bass, we
report frame F1 and note F1 with a 50-ms onset tolerance, together
with their macro-average. VA Rate is defined as
$\mathrm{F1}_{\mathrm{avg}}/\mathrm{F1}_{\mathrm{agn}}\times100\%$.
All systems use pitch-transposition augmented training data.
$^{\star}$ denotes baselines adapted for part-aware transcription.}
\vspace{1mm}
\label{tab:youchorale_results_part}
\centering
\small
\setlength{\tabcolsep}{1.6pt}
\renewcommand{\arraystretch}{0.88}
\setlength{\aboverulesep}{0.25ex}
\setlength{\belowrulesep}{0.25ex}

\begin{tabularx}{\textwidth}{
@{}l *{12}{>{\centering\arraybackslash}X}@{}
}
\toprule
\multirow{2}{*}{Model}
& \multicolumn{2}{c}{Soprano}
& \multicolumn{2}{c}{Alto}
& \multicolumn{2}{c}{Tenor}
& \multicolumn{2}{c}{Bass}
& \multicolumn{2}{c}{Average}
& \multicolumn{2}{c}{VA Rate (\%)} \\
\cmidrule(lr){2-3}
\cmidrule(lr){4-5}
\cmidrule(lr){6-7}
\cmidrule(lr){8-9}
\cmidrule(lr){10-11}
\cmidrule(lr){12-13}
& Frame & Note
& Frame & Note
& Frame & Note
& Frame & Note
& Frame & Note
& Frame & Note \\
\midrule

YourMT3+$^{\star}$~\cite{yourmt3_plus}
& 0.371 & 0.083
& 0.297 & 0.068
& 0.294 & 0.073
& 0.371 & 0.093
& 0.333 & 0.079
& 72.87 & 62.20 \\

Yu et al.$^{\star}$~\cite{yu2024note}
& 0.528 & 0.183
& \underline{0.451} & 0.151
& 0.436 & 0.138
& 0.560 & 0.186
& 0.494 & 0.165
& 76.35 & 69.62 \\
\midrule

\rowcolor{OursBlue}
PagCT + Post-VA
& 0.501 & 0.180
& 0.439 & 0.145
& 0.447 & 0.156
& 0.568 & 0.219
& 0.489 & 0.175
& 66.26 & 45.81 \\

\rowcolor{OursBlue}
PawCT w/o union-loss
& 0.417 & 0.192
& 0.338 & 0.169
& 0.343 & 0.178
& 0.419 & 0.222
& 0.379 & 0.190
& 51.36 & 49.74 \\

\rowcolor{OursBlue}
PawCT
& 0.530 & \underline{0.228}
& 0.430 & \underline{0.191}
& 0.429 & \underline{0.197}
& 0.552 & \underline{0.252}
& 0.485 & \underline{0.217}
& 65.72 & 56.81 \\

\rowcolor{OursBlue}
PawCT-RP
& \textbf{0.542} & 0.210
& \textbf{0.453} & 0.185
& \textbf{0.458} & 0.194
& \textbf{0.585} & 0.247
& \textbf{0.510} & 0.209
& 69.11 & 54.71 \\

\rowcolor{OursBlue}
PawCT-RP-OC
& \underline{0.535} & \textbf{0.231}
& 0.447 & \textbf{0.197}
& \underline{0.451} & \textbf{0.208}
& \underline{0.578} & \textbf{0.264}
& \underline{0.503} & \textbf{0.225}
& 68.16 & 58.90 \\
\bottomrule
\end{tabularx}
\vspace{-5mm}
\end{table*}

%% file: latex/tables/cross_dataset.tex
\begin{table}[t]
\setlength{\belowcaptionskip}{2pt}
\caption{Cross-dataset evaluation of part-aware note F1 with a 50-ms onset tolerance on CSD and Cantor\'ia. Models are trained only on YouChorale. Bold denotes the best result.}
\vspace{1mm}
\label{tab:cross_dataset_results}
\centering
\fontsize{9}{11}\selectfont
\setlength{\tabcolsep}{1.8pt}
\renewcommand{\arraystretch}{0.92}
\begin{tabular}{@{}llccccc@{}}
\toprule
Dataset & Model & Soprano & Alto & Tenor & Bass & Avg. \\
\midrule
\multirow{3}{*}{\textbf{CSD}}
& PagCT + Post-VA
& 0.073 & 0.122 & 0.131 & 0.143 & 0.118 \\
& PawCT
& 0.140 & 0.175 & 0.183 & \textbf{0.145} & 0.161 \\
& PawCT-RP-OC
& \textbf{0.145} & \textbf{0.223} & \textbf{0.199}
& 0.130 & \textbf{0.174} \\
\midrule
\multirow{3}{*}{\textbf{Cantor\'ia}}
& PagCT + Post-VA
& 0.179 & 0.115 & 0.085 & 0.084 & 0.115 \\
& PawCT
& 0.240 & 0.239 & 0.275 & 0.328 & 0.271 \\
& PawCT-RP-OC
& \textbf{0.329} & \textbf{0.273} & \textbf{0.300}
& \textbf{0.348} & \textbf{0.313} \\
\bottomrule
\end{tabular}
\vspace{-5mm}
\end{table}